\documentclass[preprint,12pt]{elsarticle}

\usepackage{algpseudocode}
\usepackage{amssymb}
\usepackage{stmaryrd}
\usepackage{amsmath}
\usepackage{amsthm}
\theoremstyle{definition}
\newtheorem{definition}{Definition}[section]
\theoremstyle{lemma}
\newtheorem{lemma}{Lemma}[section]
\theoremstyle{theorem}
\newtheorem{theorem}{Theorem}[section]
\newcommand{\ignoreme}[1]{}

\begin{document}

\begin{frontmatter}



\title{Partial Predicate Abstraction and Counter-Example Guided Refinement}


\author{Tuba Yavuz}

\address{}

\begin{abstract}
In this paper we present a counter-example guided abstraction and approximation  refinement (CEGAAR) technique for {\em partial predicate abstraction}, which  
 combines  predicate abstraction and fixpoint approximations for model checking infinite-state systems. The proposed approach incrementally considers growing sets of predicates for abstraction refinement. The novelty of the approach stems from recognizing source of the imprecision: abstraction or approximation. We use Craig interpolation to deal with imprecision due to abstraction. In the case of imprecision due to approximation, we delay application of the approximation. Our experimental results on a variety of models provide insights into effectiveness of partial predicate abstraction as well as refinement techniques in this context. 
\end{abstract}

\begin{keyword}


Model checking \sep Predicate Abstraction \sep Fixpoint Approximations
\end{keyword}

\end{frontmatter}


\section{Introduction}

State-explosion is an inherent problem in model checking. Every model checking tool - no matter how optimized  - will report or demonstrate one of the following for systems that push its limits: out of memory error, non-convergence, or inconclusive result. As the target systems of interest (hardware, software, or biological systems) grow in terms of complexity, and consequently in size, a great deal of manual effort is spent on verification engineering to produce usable results. We admit that this effort will always be needed. However, we also think that hybrid approaches should be employed to push the limits for automated verification.

Abstract interpretation framework \cite{CC77} provides a theoretical basis for sound verification of finite as well as infinite-state systems. Two major elements of this framework are abstraction and approximation. Abstraction defines a mapping between a concrete domain and an abstract domain (less precise) in a conservative way so that when a  property is satisfied for an abstract state the property also holds for the concrete states that map to the abstract state. 
Approximation, on the other hand, works on values in the same domain and provides a lower or an upper bound.  Abstraction is a way to deal with the state-explosion problem whereas approximation is a way to achieve convergence and hence potentially a conclusive result. When an infinite-state system is considered there are three basic approaches that can be employed: pure abstraction, pure approximation\footnote{Assuming the logic that describes the system is decidable.}, and a combination of abstraction and approximation. 

The most popular abstraction technique is predicate abstraction \cite{GS97}, in which the abstract domain consists of a combination of valuations of Boolean variables that represent truth values of a fixed set of predicates on the variables from the concrete system. Since it is difficult to come up with the right set of predicates that would yield a precise analysis, predicate abstraction has been combined with the counter-example guided abstraction refinement (CEGAR) framework \cite{CGJ00}. Predicate abstraction requires computing a quantifier-free version of the transformed system and, hence, potentially involves an exponential number of queries to the underlying SMT solver. 

A widely used approximation technique is {\em widening}. The widening operator takes two states belonging to the same domain and computes an over-approximation of the two. A key point of the widening operator is the guarantee for stabilizing an increasing chain after a finite number of steps. So one can apply the widening operator to the iterates of a non-converging fixpoint computation and achieve convergence, where the last iterate is an over-approximation of the actual fixpoint.  
In this paper we use   
an implementation of the widening operator for convex polyhedra \cite{CH78} that is used in the infinite-state model checker Action Language Verifier (ALV) \cite{YB09}.  ALV uses fixpoint approximations to check whether a CTL property is satisfied by an infinite-state system \cite{BGP97}.  

In \cite{Yav16} we introduced {\em partial predicate abstraction} that  combines predicate abstraction with widening 
for infinite-state systems described in terms of Presburger arithmetic. In partial predicate abstraction only the variables that are involved in the predicates are abstracted and all other variables are preserved in their concrete domains. In this paper, we present a counter-example guided abstraction and approximation refinement ({\em CEGAAR}) technique to deal with cases where the initial set of predicates are not precise enough to provide a conclusive result. The novelty of the approach stems from the fact that it can identify whether an infeasible counter-example is generated due to imprecision of the abstraction or imprecision of the approximation. Once the type of imprecision is identified, it uses the appropriate refinement. To refine the abstraction, it computes a Craig interpolant \cite{Cra57} for the divergence point. To refine the approximation, it delays the widening for least fixpoint computations and increases the number of steps for greatest fixpoint computations. 
 We implemented the combined approach by extending the Action Language Verifier (ALV) \cite{YB09} with the CEGAAR technique.
 Our experimental results show that approximation and abstraction refinement can be merged in an effective way. 
 
The rest of the paper is organized as follows. We first present the basic definitions and key results of the two approaches, approximate fixpoint computations and predicate abstraction in the context of CTL model checking, in Section \ref{sec:prel}.  
Section \ref{sec:approach} presents the partial predicate abstraction approach and demonstrates soundness of combining the two techniques. 
Section \ref{sec:cegaar} presents the core algorithms for the CEGAAR technique. 
 Section \ref{sec:exp} presents the experimental results. Section \ref{sec:relwork} discusses related work and Section \ref{sec:conc} concludes with directions for future work.

\section{Preliminaries}
\label{sec:prel}

In this paper, we consider transition systems that are described in terms of boolean and unbounded integer variables. 

\begin{definition}
An infinite-state transition system is described by a Kripke structure $T=(S, I, R, V)$, where $S$, $I$, $R$, and $V$ denote the state space, set of initial states, the transition relation, and the set of state variables, respectively. $V=V_{bool} \cup V_{int}$ such that $S \subseteq \mathcal{B}^{|V_{bool}|} \times \mathcal{Z}^{|V_{int}|}$, $I \subseteq S$, and $R \subseteq S \times S$.
\end{definition}

\begin{definition}
Given a Kripke structure, $T=(S, I, R, V)$ and a set of states $A \subseteq S$, the post-image operator, 
$post[R](A)$, computes the set of states that can be reached from the states in $A$ in one step: 
\[post[R](A)=\{ b \ | \ a \in A \ \wedge \ (a,b) \in R \}.\]
Similarly, the pre-image operator, $pre[R](A)$, computes the set of states that can reach the states in $A$  in one step: 
\[pre[R](A)=\{ b \ | \ a \in A \ \wedge \ (b,a) \in R \}.\]
\end{definition}

\paragraph{Model Checking via Fixpoint Approximations.} 
Symbolic Computation-Tree Logic (CTL) model checking algorithms decide whether a given Kripke structure, $T=(S, I, R, V)$, satisfies a given CTL correctness property, $f$, by checking whether $I \subseteq \llbracket f \rrbracket_T$, where $\llbracket f \rrbracket_T$ denotes the set of states that satisfy $f$ in $T$. Most CTL operators have either least fixpoint ($EU$, $AU$) or greatest fixpoint ($EG$, $AG$) characterizations in terms of the pre-image operator. 

\begin{figure}[th!]
\centering
\begin{footnotesize}
\begin{tabular}{cl} \hline &  \\
Variables & $s$, $t$, $a_1$, $a_2$, $z$: integer \\
                      & $pc1$, $pc2$: {think, try, cs} \\ 
Initial State: & $s=t \wedge pc_1=think \wedge pc_2=think$ \\ 
Transitions: &  \\ 
  $r^{try}_i $ & $\equiv pc_i=think \wedge a_i'= t \wedge t'=t+1 \wedge pc_i'=try$ \\
  $r^{cs}_i $ & $\equiv pc_i=try \wedge s \geq a_i \wedge z'=z+1 \wedge pc_i'=cs $\\
  $r^{think}_i $ & $\equiv pc_i=cs \wedge s'=s+1 \wedge z'=z-1 \wedge pc_i'=think$ \\
Transition Relation: & $\bigvee_{i=1,2} r^{try}_i \vee r^{cs}_i \vee r^{think}_i$ \\ & \\ \hline 
\end{tabular}
\end{footnotesize} \vspace{0.5mm}
\caption{The ticket mutual exclusion algorithm for two processes. Variable $z$ is an addition to demonstrate the merits of the proposed approach.}
\label{fig:ticket}
\end{figure}

Symbolic CTL model checking for infinite-state systems may not converge. Consider the so-called ticket mutual exclusion model for two processes \cite{And91} given in Figure \ref{fig:ticket}. Each process gets a ticket number before attempting to enter the critical section. There are two global integer variables, $t$ and $s$, that show the next ticket value that will be available to obtain and the upper bound for tickets that are eligible to enter the critical section, respectively. Local variable $a_i$ represents the ticket value held by process $i$. We added variable $z$ to model an update in the critical region. It turns out that checking $AG(z\leq1)$ for this model does not terminate. 

One way is to compute an over or an under approximation to the fixpoint computations as proposed in \cite{BGP97} and check $I \subseteq \llbracket f \rrbracket^{-}_T$, i.e., check whether all initial states in $T$ satisfy an under-approximation (denoted by superscript $-$) of the correctness property or check $I \cap \llbracket \neg f \rrbracket^{+}_T\not=\emptyset$, i.e., check whether no initial state satisfies an over-approximation of the negated correctness property. If so, the model checker certifies that the property is satisfied. Otherwise, no conclusions can be made without further analysis.

The key in approximating a fixpoint computation is the availability of over-approximating and under-approximating operators. So we give the basic definitions and a brief explanation here and refer the reader to \cite{CH78,BGP97} for technical details on the implementation of these operators for Presburger arithmetic.

\begin{definition}
Given a complete lattice $(L, \sqsubseteq, \sqcap, \sqcup, \perp, \top)$,  
$\triangle: L \times L \to L$,  is a widening operator iff 
\begin{itemize}
\item $\forall x, y \in L.  \ x \sqcup y \sqsubseteq x \triangle y$,
\item For all increasing chains $x_0 \sqsubseteq x_1 \sqsubseteq  ... x_n$ in L, the increasing chain 
$y_0=x_0, ..., y_{n+1} = y_n \triangle x_{n+1}, ...$ is not strictly increasing, i.e., stabilizes after a number of terms.  
\end{itemize}
\end{definition}

\begin{definition}
Given a complete lattice $(L, \sqsubseteq, \sqcap, \sqcup, \perp, \top)$,  
$\nabla: L \times L \to L$,  is a dual of the widening operator iff 
\begin{itemize}
\item $\forall x, y \in L.  x \nabla y  \sqsubseteq x \sqcap y$,
\item For all decreasing chains $x_0 \sqsupseteq x_1 \sqsupseteq  ... x_n$ in L, the decreasing chain 
$y_0=x_0, ..., y_{n+1} = y_n \nabla x_{n+1}, ...$ is not strictly decreasing, i.e., stabilizes after a number of terms.  
\end{itemize}
\end{definition}

The approximation of individual temporal operators in a CTL formula is decided recursively based on the type of approximation to be achieved and whether the operator is preceded by a negation. The over-approximation can be computed using the widening operator for least fixpoint characterizations and terminating the fixpoint iteration after a finite number of steps for greatest fixpoint characterizations. The under-approximation can be computed using the dual of the widening operator for the greatest  fixpoint characterizations and terminating the fixpoint iteration after a finite number of steps for the least fixpoint characterizations. Another heuristic that is used in approximate symbolic model checking is to compute an over-approximation (denoted by superscript $+$) of the set of reachable states ($(\mu Z. I \vee post[R](Z))^{+}$), a least fixpoint characterization, and to restrict all the fixpoint computations within this set.

\begin{lemma}
\label{lemma:appr}
Given an infinite-state transition system $T=(S,I,R,V)$ and $T^+=((\mu Z. I \vee post[R](Z))^{+}, I, R, V)$, and a temporal property $f$, the conclusive results obtained using fixpoint approximations for the temporal operators and the approximate set of reachable states are sound, i.e., $(I \subseteq \llbracket f \rrbracket^{-}_{T^+} \ \vee \ I \cap \llbracket \neg f \rrbracket^{+}_{T^+} = \emptyset) \to \ T \models f$ (see \cite{BGP97} for the proof).
\end{lemma}

So for the example model in Figure  \ref{fig:ticket},  an over-approximation to $EF(z>1)$, the negation of the correctness property, is computed using the widening operator.  Based on the implementation of the widening operator in \cite{YB09}, it turns out that the initial states do not intersect with $\llbracket EF(z>1) \rrbracket^+_{ticket2}$ and hence the model satisfies $AG(z\leq1)$.

\paragraph{Abstract Model Checking and Predicate Abstraction.}

\begin{definition}
Let $\varphi$ denote a set of predicates over integer variables. Let $\varphi_i$ denote a predicate in $\varphi$ and $b_i$ denote the 
boolean variable that corresponds to $\varphi_i$. $\bar{\varphi}$ represents an ordered sequence (from index 1 to $|\varphi|$) of predicates in $\varphi$. The set of variables that appear in $\varphi$ is denoted by $V(\varphi)$. Let $\varphi'$ denote the set of next state predicates obtained from $\varphi$ by replacing variables in each predicate $\varphi_i$ with their primed versions. Let $b$ denote the set of $b_i$ that corresponds to each $\varphi_i$. Let $V_{\sharp} = V_{\natural} \cup b \setminus V(\varphi)$, where $V_{\natural}$ denotes the set of variables in the concrete model.  
\end{definition}

\paragraph{Abstracting states.} A concrete state $s^{\natural}$ is predicate abstracted using a mapping function $\alpha$ via a set of predicates $\varphi$ by introducing a predicate boolean variable $b_i$ that represents  predicate $\varphi_i$ and existentially quantifying the concrete variables $V(\varphi)$ that appear in the predicates:

\begin{equation}
\label{eq:alpha}
\alpha(s^{\natural}) = \exists V(\varphi). (s^{\natural} \ \wedge \ \bigwedge^{|\varphi|}_{i=1} \varphi_i \iff b_i ).
\end{equation}

\paragraph{Concretization of abstract states.} An abstract state $s^{\sharp}$ is mapped back to all the concrete states it represents by replacing each predicate boolean variable $b_i$ with the corresponding predicate $\varphi_i$:
\begin{equation}
\label{eq:gamma}
\gamma(s^{\sharp}) = s^{\sharp} [\bar{\varphi}/\bar{b}]
\end{equation}

Abstraction function $\alpha$ provides a safe approximation for states:

\begin{lemma}
\label{lemma:galoisState}
$(\alpha, \gamma)$, as defined in Equations \ref{eq:alpha} and \ref{eq:gamma}, defines a Galois connection, i.e., $\alpha$ and $\gamma$ are monotonic functions and $s^{\natural} \subseteq \gamma(\alpha(s^{\natural}))$ and $\alpha(\gamma(s^{\sharp}))=s^{\sharp}$ (see the Appendix for the proof).
\end{lemma}

A concrete transition system can be conservatively approximated by an abstract transition system through a simulation relation  or a surjective mapping function involving the respective state spaces:

\begin{definition} (Existential Abstraction)
\label{def:exabs}
Given transition systems $T_1=(S_1, I_1, R_1, V_1)$ and $T_2=(S_2,I_2,R_2,V_2)$, $T_2$ approximates $T_1$ (denoted $T_1 \sqsubseteq_{h} T_2$) iff
\begin{itemize}
\item $\exists s_1. (h(s_1) = s_2 \ \wedge \ s_1 \in I_1)$ implies $s_2 \in I_2$,
\item $\exists s_1, s_1'. (h(s_1)=s_2 \ \wedge \ h(s_1')=s_2' \ \wedge \ (s_1,s_1') \in R_1)$ implies $(s_2, s_2') \in R_2$,
\end{itemize} 
where $h$ is a surjective function from $S_1$ to $S_2$.
\end{definition}

It is a known  \cite{Loi95} fact that one can use a Galois connection $(\alpha, \gamma)$ to construct an approximate transition system. Basically, $\alpha$ is used as the mapping function and $\gamma$ is used to map properties of the approximate or abstracted system to the concrete system:

\begin{definition}
\label{def:concFormula}
Given transition systems $T_1=(S_1, I_1, R_1, V_1)$ and $T_2=(S_2,I_2,$\\$R_2,V_2)$,
assume that $T_1 \sqsubseteq_{\alpha} T_2$, the ACTL formula $\phi$ describes properties of $T_2$, and $(\alpha,\gamma)$ forms a Galois connection.  
 $C(\phi)$ represents a transformation on $\phi$ that descends on the subformulas recursively and transforms every atomic atomic formula $a$ with $\gamma(a)$ (see \cite{CGL94} for details).
\end{definition}

For example, let $\phi$ be $AG(b_1 \vee b_2)$, where $b_1$ and $b_2$ represent $z=1$ and $z<1$, respectively,  when the model in Figure \ref{fig:ticket} is predicate abstracted wrt to the set of predicates $\varphi=\{z=1, z<1\}$ and the Galois connection  $(\alpha, \gamma)$ defined as in Equations \ref{eq:alpha} and \ref{eq:gamma}. Then, $C(\phi)=AG(z\leq1)$.

The preservation of ACTL properties when going from the approximate system to the concrete system is proved for existential abstraction in \cite{CGL94}. Here, we adapt it to an instantiation of existential abstraction using predicate abstraction as in \cite{CGT03}:

\begin{lemma}
\label{lemma:amc}
Assume $T_1 \sqsubseteq_{\alpha} T_2$, $\phi$ denotes an ACTL formula that describes a property of $T_2$, $C(\phi)$ denotes the transformation of the correctness property as in Definition \ref{def:concFormula}, and $(\alpha,\gamma)$ forms a Galois connection and defines predicate abstraction and concretization as given in Equations \ref{eq:alpha} and \ref{eq:gamma}, respectively. Then, $T_2 \models \phi$ implies $T_1 \models C(\phi)$.
\begin{proof}
{\em Preservation of atomic properties:}
If a state $s_2$ in $T_2$ satisfies an atomic abstract property $\phi$, due to the correctness preserving 
property of a Galois connection, $s_2$ also satisfies $\gamma(\phi)$ \cite{Niel99}. Due to soundness of the mapping between the states in $T_1$ to states in $T_2$ and monotonic property of $\alpha$ and $\gamma$, any state $s_1$ in $T_1$ that gets mapped to $s_2$, that is every state in $\gamma(s_2)$ also satisfies $\gamma(\phi)$.\\ 
{\em Preservation of ACTL Properties:} 
Follows from Corollary 1 in \cite{CGL94} and using 
$\alpha$ as the mapping function $h$ in \cite{CGL94}. 
\end{proof} 
\end{lemma}

\section{Partial Predicate Abstraction}
\label{sec:approach}

In Section \ref{sec:partial}, we introduce a symbolic abstraction operator for transitions and an over-approximating abstract post operator derived from it.  The abstract post operator enables partial predicate abstraction of an infinite-state system. Section \ref{sec:comb} elaborates on 
the proposed hybrid approach that combines predicate abstraction and fixpoint approximations to perform CTL model checking of infinite-state systems. It also demonstrates soundness of the hybrid approach, which follows from the soundness results of the individual approaches and the over-approximating nature of the proposed abstract post operator.

\subsection{Computing A Partially Predicate Abstracted Transition System}
\label{sec:partial}

We compute an abstraction of a given transition system via a set of predicates such that only the variables that appear in the predicates disappear, i.e., existentially quantified, and all the other variables are preserved in their concrete domains and in the exact semantics from the original system. As an example, using the set of predicates $\{z=1, z<1\}$, we can partially abstract the model in Figure \ref{fig:ticket} in a way that $z$ is removed from the model, two new boolean variables $b_1$ (for $z=1$) and $b_2$ (for $z<1$) are introduced, and $s$, $t$, $a_1$, $a_2$, $pc_1$, and $pc_2$ remain the same as in the original model.

\paragraph{Abstracting transitions.} A concrete transition $r^{\natural}$ is predicate abstracted using a mapping function $\alpha^{\tau}$ via a 
set of current state  predicates $\varphi$ and a set of next state  predicates $\varphi'$  by introducing a predicate boolean variable $b_i$ that represents predicate $\varphi_i$ in the current state and a predicate boolean variable $b_i'$ that represents predicate $\varphi_i$ in the next state and existentially quantifying the current and next state concrete variables $V(\varphi) \cup V(\varphi')$ that appear in the current state and next state predicates:

\begin{equation}
\label{eq:alphaT}
\alpha^\tau(r^{\natural}) = \exists V(\varphi). \exists V(\varphi'). (r^{\natural} \ \wedge \ CS \ \wedge \ \bigwedge^{|\varphi|}_{i=1} \varphi_i \iff b_i \wedge \ \bigwedge^{|\varphi|}_{i=1} \varphi_i' \iff b_i'  ),
\end{equation}

where $CS$ represents a consistency constraint that if all the abstracted variables that appear in a predicate remains the same in the next state then the corresponding boolean variable is kept the same in the next state:

\[CS = \bigwedge_{\varphi_i \in \varphi} ((\bigwedge_{v \in V(\varphi_i)} v'=v ) \implies b_i' \iff b_i).\] 

\paragraph{Concretization of abstract transitions.} An abstract transition $r^{\sharp}$ is mapped back to all the concrete transitions it represents by replacing each current state boolean variable $b_i$ with the corresponding current state predicate $\varphi_i$ and each next state boolean variable $b_i'$ with the corresponding next state predicate $\varphi_i'$:

\[\gamma^{\tau}(r^{\sharp}) = r^{\sharp} [\bar{\varphi},\bar{\varphi}'/\bar{b},\bar{b}']\]

For instance, for the model in Figure \ref{fig:ticket} and predicate set $\phi=\{z=1, z<1\}$, partial predicate abstraction of $r^{cs}_i$, $\alpha^{\tau}(r^{cs}_i)$, is  computed as 

\begin{equation}
\begin{split}
pc_i= & try \ \wedge \ s \geq a_i \ \wedge \ ((b_1 \wedge \neg b_2  \wedge \neg b'_1  \wedge  \neg b'_2)  \vee \ (\neg b_1 \wedge  b_2 \wedge  (b'_1  \vee  b'_2)) \\
& \vee \ (\neg b_1 \wedge \neg b_2 \wedge \neg b'_1 \wedge \neg b'_2))   \ \wedge \ pc_i'=cs .
\end{split}
\end{equation}

It is important to note that the concrete semantics pertaining to the integer variables $s$ and $a_i$ and the enumerated variable $pc_i$ are preserved in the partially abstract system.

Abstraction function $\alpha^{\tau}$ represents a safe approximation for transitions:

\begin{lemma}
\label{lemma:galoisTrans}
$(\alpha^{\tau}, \gamma^{\tau})$ defines a Galois connection (see the Appendix for the proof). 
\end{lemma}

One can compute an over-approximation to the set of reachable states via an over-approximating abstract post operator that computes the abstract successor states:

\begin{lemma}
\label{lemma:appPost}
$\alpha^{\tau}$ provides an over-approximate post operator:
\[ post[r^{\natural}](\gamma(s^{\sharp})) \ \subseteq \   \gamma(post[\alpha^{\tau}(r^{\natural})](s^{\sharp}))   \]

\begin{proof}
\begin{equation}
\begin{split}
post[\tau^{\natural}](\gamma(s^{\sharp})) \ \subseteq & \ post[\gamma^{\tau}(\alpha^{\tau}(\tau^{\natural}))](\gamma(s^{\sharp})) (\mbox{due to Lemma \ref{lemma:galoisTrans}}) \label{step1} \\
\end{split}
\end{equation}
We need to show the following:
\begin{equation}
\begin{split}
post[\gamma^{\tau}(\alpha^{\tau}(\tau^{\natural}))](\gamma(s^{\sharp})) \ & \subseteq \ \gamma(post[\alpha^{\tau}(\tau^{\natural})](s^{\sharp}))\ \label{step2} \\
post[\gamma^{\tau}(\tau^{\sharp})](\gamma(s^{\sharp})) \ & \subseteq \ \gamma(post[\tau^{\sharp}](s^{\sharp}))\ \\
(\exists V_{\natural}. \ \tau^{\sharp}[\bar{\varphi},\bar{\varphi}'/\bar{b},\bar{b}'] \ \wedge \ s^{\sharp}[\bar{\varphi}/\bar{b}])[V_{\natural}/V_{\natural}'] \ & \subseteq \ (\exists V_{\sharp}. \ \tau^{\sharp} \ \wedge \ s^{\sharp})[V_{\sharp}/V_{\sharp}'][\bar{\varphi}/\bar{b}]  \\
(\exists V_{\natural}. \ \tau^{\sharp}[\bar{\varphi},\bar{\varphi}'/\bar{b},\bar{b}'] \ \wedge \ s^{\sharp}[\bar{\varphi}/\bar{b}])[V_{\natural}/V_{\natural}'] \ & \subseteq \ (\exists V_{\sharp}. \ \tau^{\sharp} \ \wedge \ s^{\sharp})[\bar{\varphi}'/\bar{b}'][V_{\natural}/V_{\natural}']  \\
(\exists V_{\natural}. (\ \tau^{\sharp} \ \wedge \ s^{\sharp})[\bar{\varphi},\bar{\varphi}'/\bar{b},\bar{b}'])
[V_{\natural}/V_{\natural}'] \ & \subseteq \ (\exists V_{\sharp}. \ \tau^{\sharp} \ \wedge \ s^{\sharp})[\bar{\varphi}'/\bar{b}'][V_{\natural}/V_{\natural}']  \\
\end{split}
\end{equation}

\begin{equation}
\begin{split}
post[\tau^{\natural}](\gamma(s^{\sharp})) \ \subseteq & \  \gamma(post[\alpha^{\tau}(\tau^{\natural})](s^{\sharp})) (\mbox{due to Equations \ref{step1} \&  \ref{step2}}) \\
\end{split}
\end{equation}

\end{proof}
\end{lemma}

\ignoreme{
\begin{lemma}
\label{lemma:appPre}
$\alpha^{\tau}$ provides an over-approximate pre operator:
\[ pre[\tau^{\natural}](\gamma(s^{\sharp})) \ \subseteq \   \gamma(pre[\alpha^{\tau}(\tau^{\natural})](s^{\sharp}))   \]

\begin{proof}
The proof is similar to the proof of Lemma \ref{lemma:appPost}.
\end{proof}
\end{lemma}
}

 \subsection{Combining Predicate Abstraction with Fixpoint Approximations} 
 \label{sec:comb}

At the heart of the hybrid approach is a partially predicate abstracted transition system and 
we are ready to provide a formal definition:

\begin{definition}
\label{def:abstractingKripke}
Given a concrete infinite-state transition system $T^{\natural}=(S^{\natural}, I^{\natural}, R^{\natural}, V^{\natural})$  and a set of predicates $\varphi$, where $V(\varphi) \subseteq V^{\natural}_{int}$, the partially predicate abstracted transition system $T^{\sharp}=(S^{\sharp}, I^{\sharp}, R^{\sharp}, V^{\sharp})$ is defined as follows:
\begin{itemize}
\item $S^{\sharp} \subseteq \mathcal{B}^{|V^{\natural}_{bool}|+|\varphi|} \times \mathcal{Z}^{|V^{\natural}_{int} \setminus V(\varphi)|}$
\item $S^{\sharp} = \bigcup_{s^{\natural} \in S^{\natural}} \alpha(s^{\natural})$.
\item $I^{\sharp} = \bigcup_{is^{\natural} \in I^{\natural}} \alpha(is^{\natural})$.
\item $R^{\sharp} = \bigcup_{r^{\natural} \in R^{\natural}} 
\alpha^{\tau} (r^{\natural})$.
\end{itemize} 
\end{definition}

A partially predicate abstracted transition system $T^{\sharp}$ defined via $\alpha$ and $\alpha^{\tau}$ functions is a conservative approximation of the concrete transition system. 

\begin{lemma}
\label{lemma:sim}
Let the abstract transition system $T^{\sharp}=(S^{\sharp}, I^{\sharp}, R^{\sharp}, V^{\sharp})$  
be defined as in Definition \ref{def:abstractingKripke} with respect to the concrete  transition system $T^{\natural}=(S^{\natural}, I^{\natural}, R^{\natural}, V^{\natural})$ and the set of predicates $\varphi$.
$T^{\sharp}$  approximates $T^{\natural}$: $T^{\natural} \sqsubseteq_{\alpha} T^{\sharp}$.
\begin{proof}
It is straightforward to see, i.e., by construction, that $\exists s_1. (\alpha(s_1) = s_2 \ \wedge \ s_1 \in I^{\natural})$ implies $s_2 \in I^{\sharp}$. To show 
$\exists s_1, s_1'. (\alpha(s_1)=s_2 \ \wedge \ \alpha(s_1')=s_2' \ \wedge \ (s_1,s_1') \in R^{\natural})$ implies $(s_2, s_2') \in R^{\sharp}$, we need to show that $\exists s_1, s_1'. (\alpha(s_1)=s_2 \ \wedge \ \alpha(s_1')=s_2' \ \wedge \ s_1' \in post[R^{\natural}](s_1))$ implies 
$s_2' \in post[\alpha^{\tau}(R^{\natural})](s_2)$, which follows from Lemma \ref{lemma:appPost}: $s_1' \in \gamma(post[\alpha^{\tau}(R^{\natural})](s_2))$ and $\alpha(s_1') \in \alpha(\gamma(post[\alpha^{\tau}(R^{\natural})](s_2)))$, and hence $s_2' \in post[\alpha^{\tau}(R^{\natural})](s_2)$. 
\end{proof}
\end{lemma}

Therefore, ACTL properties verified  on $T^{\sharp}$ also holds for $T^{\natural}$:

\begin{lemma}
\label{lem:actl}
Let the abstract transition system $T^{\sharp}=(S^{\sharp}, I^{\sharp}, R^{\sharp}, V^{\sharp})$  
be defined as in Definition \ref{def:abstractingKripke} with respect to the concrete  transition system $T^{\natural}=(S^{\natural}, I^{\natural}, R^{\natural}, V^{\natural})$ and the set of predicates $\varphi$.
Given an ACTL property $f^{\sharp}$, $T^{\sharp} \models f^{\sharp} \ \rightarrow T^{\natural} \models \gamma(f^{\sharp})$.
\begin{proof}
Follows from Lemmas  \ref{lemma:amc} and \ref{lemma:sim}.
\end{proof}
\end{lemma}

Using fixpoint approximation techniques on an infinite-state partially predicate abstracted transition system in symbolic model checking of CTL properties  \cite{BGP97} preserves the verified ACTL properties due to Lemma \ref{lemma:appr} and Lemma \ref{lem:actl}.
 
Restricting the state space of an abstract transition system 
 $T^{\sharp}=(S^{\sharp}, I^{\sharp}, R^{\sharp})$
  with an over-approximation of the set of reachable states $T^{\sharp}_{RS}=(\mu Z. post[R^{\sharp}](Z) \ \vee \ I^{\sharp})^+, I^{\sharp}, R^{\sharp})$ also preserves the verified ACTL properties:
 
 \begin{theorem}
 \label{theorem:hybrid}
 Let the abstract transition system $T^{\sharp}=$ $(S^{\sharp},$$I^{\sharp},$$ R^{\sharp}, V^{\sharp})$  be defined as in Definition \ref{def:abstractingKripke} with respect to the concrete  transition system $T^{\natural}=(S^{\natural}, I^{\natural}, R^{\natural}, V^{\natural})$. Let $T^{\sharp}_{RS}=$ $((\mu Z. I^{\sharp} \ \vee \ post[R^{\sharp}](Z))^+$$, I^{\sharp},$$ R^{\sharp}, V^{\sharp})$. Given an ACTL property $f^{\sharp}$,   $I^{\sharp} \subseteq \llbracket f^{\sharp} \rrbracket^{-}_{T^{\sharp}_{RS}} \rightarrow T^{\natural} \models \gamma(f^{\sharp})$.
\begin{proof}
Follows from Lemma \ref{lemma:appr} that approximate symbolic model checking is sound, i.e., $I^{\sharp} \subseteq \llbracket f^{\sharp} \rrbracket^{-}_{T^{\sharp}_{RS}}$ implies $T^{\sharp} \models  f^{\sharp}$, and from Lemma \ref{lem:actl} that ACTL properties verified on the partially predicate abstracted transition system holds for the concrete transition system, i.e., $T^{\sharp} \models  f^{\sharp}$ implies $T^{\natural} \models \gamma(f^{\sharp})$.
\end{proof}
 \end{theorem}
 
 As an example, using the proposed hybrid approach one can show that the concrete model, $T^{\natural}_{ticket2}$ given in Figure \ref{fig:ticket} satisfies the correctness property $AG(z\leq1)$ by 
 first generating a partially predicate abstracted model, $T^{\sharp}_{ticket2}$, wrt the predicate set $\{z=1, z<1\}$ and performing approximate fixpoint computations to prove $AG(b_1 \vee b_2)$. Due to Theorem \ref{theorem:hybrid}, 
 if $T^{\sharp}_{ticket2, RS}$ satisfies $AG(b_1 \vee b_2)$,  it can be concluded that $T^{\natural}_{ticket2}$ satisfies $AG(z\leq1)$. 
 
 The main merit of the proposed approach is to combat the state explosion problem in the verification of problem instances  for which predicate abstraction does not provide the necessary precision (even in the case of being embedded in a CEGAR loop) to achieve a conclusive result.  In such cases  approximate fixpoint computations may turn out to be more precise. The hybrid approach may provide both the necessary precision to achieve a conclusive result and an improved performance by predicate abstracting the variables that do not require fixpoint approximations. 

\section{Counter-Example Guided Abstraction and Approximation Refinement}
\label{sec:cegaar}

\begin{figure}[th!]
\begin{footnotesize}
\begin{algorithmic}[1]
\State PreCondition: $f = \gamma(\alpha_{seed}(f)) \wedge \forall v : \mathcal{Z}. v \in V(f) \rightarrow \exists \varphi_i \in seed. v \in V(\varphi_i)$
\State {\bf global} $wideningSeed: int$ \ {\bf global} $overApprBound: int$
\State {\bf CEGAAR}($T=(S,I,R)$, $f$: ACTL property, $seedPreds$: set of predicates)
\State $\varphi$: current set of predicates
\State $bound \gets overApprBound$
\State $ws \gets wideningSeed$
\State $predWorklist \gets emptyList$
\State $predWorklist.insert(seedPreds)$
\State $approxRefinement \gets false$
\While{$true$}
 \If{$approxRefinement=false$}
   \If{$predWorklist \not = emptyList$ }
     \State $\varphi \gets predWorklist.remove()$  
     \State  $ws \gets wideningSeed$ 
     \State $bound \gets overApprBound$
   \Else \  print "UNABLE TO VERIFY" ;  break 
   \EndIf
 \EndIf
 \State Let $T^{\sharp}=(S^{\sharp} , I^{\sharp}, R^{\sharp})$, $S^{\sharp} = \alpha_{\varphi}(S)$, $I^{\sharp}=\alpha_{\varphi}(I)$, $R^{\sharp}=\alpha_{\varphi}(R)$
 \State Let $iter: ECTL \to \text{list of } \mathcal{P}(S^{\sharp})$ \ $iter \gets Compute(\neg \alpha_{\varphi}(f), T^{\sharp},ws,bound)$
 \State Let $Sol \gets iter(\neg \alpha_{\varphi}(f)).last$
 \If{$S^{\sharp} \wedge Sol = false$}
    \State print "ALL INITIAL STATES ARE SATISFIED"; break
 \Else
     \State Let $Wit \gets emptyList$
     \State $(dT, dD, Conf) \gets  \text{\bf GenAbsWitness} (T, T^{\sharp}, I, I^{\sharp}, \neg \alpha_{\varphi}(f), iter, Wit)$    
     \If{$dT  = ABST$}
         \State $approxRefinement \gets false$
         \State Let $Conf = (conf_c,conf_a)$ 
         \State $\varphi' \gets CraigInterpolant(conf_c,conf_a)$
         \For{each $p \in \varphi'$}
          \State $predWorklist.insert(\varphi \cup \{p\})$     
         \EndFor 
      \ElsIf{$dT = APPR$} 
         \State $approxRefinement \gets true$
          \State $ws \gets dD + 1$;  \State $bound \gets max(overAprrBound,dD+1)$
     \Else \  print "PROPERTY VIOLATED" 
     \EndIf 
 \EndIf
\EndWhile
\end{algorithmic}
\end{footnotesize}
\caption{An algorithm for counter-example guided abstraction and approximation refinement.}
\label{alg:cegaar}
\end{figure}

In this section, we present a counter-example guided abstraction and approximation refinement technique for partial predicate abstraction (CEGAAR) in the context of CTL model checking. The individual techniques that are combined in partial predicate abstraction have their own specialized techniques for refinement. Counter-example guided abstraction refinement (CEGAR) \cite{CGJ00} has been shown to be an effective way for improving precision of predicate abstraction by inferring new predicates based on the divergence between the abstract counter-example path and the concrete paths. Approximation refinement, on the other hand, involves shrinking  the solution set for over-approximations and expanding the solution set for under-approximations. 

In partial predicate abstraction, a fundamental dilemma is whether to apply abstraction refinement or approximation refinement. Since model checking of infinite-state systems is undecidable and both CEGAR and approximation refinement techniques may not terminate and, hence, may fail to provide a conclusive result, whatever approach we follow in applying these alternative techniques may not terminate either. 

In this paper, we choose to guide the refinement process using counter-examples. However, a novel aspect of our approach is its ability to recognize  source of the imprecision. So, if  the imprecision is due to approximation, it switches from abstraction refinement to approximation refinement. After entering the approximation refinement mode, it may switch back to abstraction refinement, end with a conclusive result, or may keep staying in the same mode. This process of possibly interleaved refinement continues until the property is verified, a real counter-example is reached, or no new predicates can be inferred.

Figure \ref{alg:cegaar} shows the {\bf CEGAAR} algorithm. It gets a concrete transition system, the concrete correctness property to be checked, and a set of seed predicates. It is important that any integer variable that appears in the property can be precisely abstracted using the seed predicates so that Theorem \ref{theorem:hybrid} can be applied in a sound way. This fact is specified in the precondition of the algorithm. The algorithm also uses some global settings: the widening seed $wideningSeed$ and the over-approximation bound $overApprBound$. The former parameter decides how early in least fixpoint computations widening can be applied, e.g., 0 means starting from the first iteration, and the latter parameter decides when to stop the greatest fixpoint computation. Stopping early results in a less precise approximation than stopping at a later stage. However, the overhead gets bigger as the stopping is delayed. The algorithm keeps local variables $ws$ and $bound$ that receive their initial values from the global variables $wideningSeed$ and $overApprBound$, respectively (lines 5-6).

The algorithm keeps a worklist, which is a list of set of predicates to be tried for partial predicate abstraction. A challenge in CEGAR is the blow-up in the number of predicates as new predicates get inferred. To deal with this problem, we have used a breadth-first search (BFS) strategy to explore the predicate choices that are stored in the worklist until  a predicate set  producing a conclusive result can be found. The work list is initialized to have one item, the seed set of predicates, before the main loop gets started (lines 7-8). 

The algorithm runs a main loop starting from an initial state, where abstraction is the current refinement strategy (line 9). In abstraction refinement mode, the algorithm
 removes a predicate set from the work list to use as the current predicate set $\varphi$, resets approximation parameters to their default global values (lines 13-15),  and computes an abstract version of the transition system via partial predicate abstraction with the current predicate set (line 19).  Then (line 20) the algorithm computes the fixpoint for the negation of the property, which happens to be an ECTL\footnote{Fragment of CTL in which only existential versions of temporal operators appear.} property. The fixpoint iterates are stored in a map (declared in line 20), which can be queried for the parts of the formula to get the relevant fixpoint iterates, which are stored in a list. As an example, we can access the fixpoint solution for subformula $f_1$ with $iter(f_1).last()$ expression. Indices start at 1 and both $iter(f_1)[1]$ and $iter(f_1).first()$ represent the first iterate and both $iter(f_1).last()$ and $iter(f_1).[iter(f_1).size()]$ represent the solution set. 
 
 If the fixpoint solution to the negation of the property does not have any states in common with the set of initial states, then the property is satisfied and the algorithm terminates (lines 22-23). Otherwise, the property is not satisfied in the abstract system. At this point, {\bf GenAbsWitness} is called to find out if there is a divergence between the abstract witness path for the negation of the property and the concrete transition system. If there is a divergence due to abstraction (line 27), refinement mode will set to abstraction refinement (line 28) and a Craig Interpolant is computed (line 30) for the reachable concrete states that can never reach the divergence point, {\em deadend} states, and the concrete states that can, {\em bad} states \cite{CTV03}. We use the half-space interpolants algorithm presented in \cite{AM13} to compute a compact refinement predicate.  We encode the constraint for the half-space interpolant using the Omega Library \cite{omega}, the polyhedra library used in ALV. If the number of variables used in the encoding reaches the limit set by the Omega Library, we revert back to a mode where we collect all the predicates in the deadend states.
  
 When a set of refinement predicates $\varphi'$ is discovered for a given spurious counter-example path, rather than extending the current predicate set $\varphi$ with $\varphi'$ in one shot, it considers as many extensions of $\varphi$ as $|\varphi'|$ by extending $\varphi$ with a single predicate from $\varphi'$ at a time (lines 31-33) and adds all these predicate sets to the queue to be explored using BFS. In the context of partial predicate abstraction, this strategy has been more effective in generating conclusive results compared to adding all refinement predicates at once, which has caused blow-ups in time and/or memory leading to inconclusive results for our benchmarks. 

If the divergence between the abstract witness path and the concrete transition system is due to approximation, refinement mode is switched to approximation refinement and approximation parameters are updated based on the depth of the abstract witness path up to the divergence point (lines 35-37). So in this mode, rather than updating the current predicate set, the same abstract transition system is used with updated approximation parameters.  The process will continue until a conclusive result is obtained or a real witness to the negation of the property can be found, i.e. the divergence is  due to neither the abstraction nor the approximation (lines 38-39).

\subsection{Divergence Detection}
\label{sec:div}

\begin{figure}[th!]
\begin{footnotesize}
\begin{algorithmic}[1]
\State {\bf GenAbsWitness}($T=(S,I,R)$, $T^{\sharp}=(S^{\sharp}, I^{\sharp}, R^{\sharp})$, $s$: $S$, $s^{\sharp}$: $S^{\sharp}$, $f: ECTL$, $iter: ECTL \to \text{list of }\mathcal{P}(S^{\sharp})$, $witness: \text{list of }\mathcal{P}(S^{\sharp})$)
\State {\em Output:} $divType : \{ABST, APPR, FAIL,INVALID\}$, $divDepth: int$, $Conf: S \times S$  
\State $witness \gets emptyList$
\State {\bf return} {\bf GenAbsWitnessHelper}$(T, T^{\sharp}, s, false, s^{\sharp}, false, f, iter, witness)$
\State {\bf GenAbsWitHelper}($T=(S,I,R)$, $T^{\sharp}=(S^{\sharp}, I^{\sharp}, R^{\sharp})$, $s$: $S$, $s_{prev}$: $S$,$s^{\sharp}$: $S^{\sharp}$, $s^{\sharp}_{prev}$: $S^{\sharp}$, $f: ECTL$, $iter: ECTL \to \text{list of }\mathcal{P}(S^{\sharp})$, $witness: \text{list of } \mathcal{P}(S^{\sharp})$)
\State {\em Output:} $divType : \{ABST, APPR, FAIL,INVALID\}$, $divDepth: int$, $Conf: S \times S$   
\State $iter_f \gets iter(f)$
\If{$f = EX f_1$} {\bf return GenAbsWit\_EX}$(T, T^{\sharp}, s, s_{prev}, s^{\sharp}, s^{\sharp}_{prev}, f, iter, witness)$
\ElsIf{$f = f_1 EU f_2$} {\bf return GenAbsWit\_EU}$(T, T^{\sharp}, s, s_{prev}, s^{\sharp}, s^{\sharp}_{prev}, f, iter, witness)$
\ElsIf{$f = EG f_1$} {\bf return GenAbsWit\_EG}$(T, T^{\sharp}, s, s_{prev}, s^{\sharp}, s^{\sharp}_{prev}, f, iter, witness)$
\ElsIf{$f = f_1 \wedge f_2$} {\bf return GenAbsWit\_AND}$(T, T^{\sharp}, s, s_{prev}, s^{\sharp}, s^{\sharp}_{prev}, f, iter, witness)$  
\ElsIf{$f=f_1 \vee f_2$}   {\bf return GenAbsWit\_OR}$(T, T^{\sharp}, s, s_{prev}, s^{\sharp}, s^{\sharp}_{prev}, f, iter, witness)$  
\Else \ {\bf return CheckValidityAndDivergence}$(s,s_{prev},s^{\sharp},s^{\sharp}_{prev},f,iter,witness)$  
\EndIf
\end{algorithmic}
\end{footnotesize}
\caption{An algorithm for generating an abstract witness up to a divergence point or up to the formula depth.}
\label{alg:witness}
\end{figure}

Divergence detection requires generation of possible witnesses to the negation of the property, which happens to be an ECTL property. Figure \ref{alg:witness} presents algorithm 
{\bf GenAbsWitness} that starts from the abstract and concrete initial states and uses the solution sets  for each subformula  stored in $iter$ to generate a witness path by adding abstract states to the list $witness$ as the solution sets are traversed. It should be noted that the witness, representing a counter-example to ACTL, is a tree-like structure as defined in \cite{CJL02}. So algorithm {\bf GenAbsWitness} traverses paths on this tree to check existence of a divergence. 
It calls algorithm {\bf GenAbsWitnessHelper}, which additionally keeps track of previous states of the abstract and concrete states.
When this algorithm is called the first time (line 4) in Figure \ref{alg:witness}, the previous abstract and the previous concrete states are passed as false in the parameter list as the current states, both in the abstract and in the concrete, represent the initial states. 
Since ECTL can be expressed with temporal operators EX, EU, EG, and logical operators $\wedge$, $\vee$, and $\neg$, we present divergence detection in Figures \ref{alg:witnessEX} - \ref{alg:witnesshelperOR} for these operators excluding $\neg$ as negation is pushed inside and appears before atomic formula only. Before explaining divergence checking for each type of operator, we first explain divergence detection algorithm.

\begin{figure}[th!]
\begin{footnotesize}
\begin{algorithmic}[1]
\State {\bf Divergence}($s$: $S$, $s_{prev}$: $S$,$s^{\sharp}$: $S^{\sharp}$, $s^{\sharp}_{prev}$: $S^{\sharp}$) 
\State {\em Output} $div$: bool, $divType : \{ABST, APPR, FAIL\}$, $Conf: S \times S$  
\State $Conf \gets (false, false)$
\If{$s_{prev}=false$ AND $s^{\sharp}_{prev}=false$} 
  $div \gets false$
\Else
\If{$s = false$}
  \State $div \gets true$
  \State $Deadend \gets s_{prev} \wedge \gamma_{\varphi}(s^{\sharp}_{prev}) \wedge \neg pre[R](\gamma_{\varphi}(s^{\sharp}))$
  \State $Bad \gets \gamma_{\varphi}(s^{\sharp}_{prev}) \wedge pre[R](\gamma_{\varphi}(s^{\sharp}))$
  \If{$Bad = false$}
      $divType \gets APPR$
  \Else 
      \State   $divType \gets ABST$
      \State $Conf \gets (Deadend, Bad)$
  \EndIf
\Else \ $div \gets false$
\EndIf
\EndIf
\State {\bf return} (div, divType, Conf)
\end{algorithmic}
\end{footnotesize}
\caption{An algorithm for checking divergence between a concrete path that ends by transitioning from $s_{prev}$ to $s$ and an abstract path that that ends by transitioning from 
$s^{\sharp}_{prev}$ to $s^{\sharp}$.}
\caption{Divergence detection on a given abstract and concrete state pair.}
\label{fig:divergence}
\end{figure}

Figure \ref{fig:divergence} presents the algorithm  {\bf Divergence} for checking divergence between an abstract path starting at an abstract initial state and ending at abstract state $s^{\sharp}$ and a parallel concrete path starting at the concrete initial path and ending at concrete state $s$. The algorithm also gets as input the previous states of $s^{\sharp}$ and $s$ as $s^{\sharp}_{prev}$ and $s_{prev}$, respectively. If both previous states are false (lines 4-5), divergence is not possible as $s^{\sharp}$ and $s$ represent the abstract and concrete initial states, respectively, and due to partial predicate abstraction being an Existential Abstraction (see \ref{def:exabs}).  
However, for any state other than the initial state, divergence may occur if  $s_{prev}$ is not enabled for the same transition that $s^{\sharp}_{prev}$ is enabled for.  In that case (line 6, holding true), the algorithm computes the {\em Deadend} states, which represent those reachable concrete states that do not have any transition to the concrete states that map to abstract state $s^{\sharp}$ (line 8), and the {\em Bad} states, which represent those concrete states that map to abstract previous state $s^{\sharp}_{prev}$ and can transition to states that map to $s^{\sharp}$ (line 9). If all the variables are predicate abstracted, the set of {\em Bad} states cannot be empty \cite{CTV03}. However, due to partial predicate abstraction and existence of fixpoint approximations for concrete integer variables, {\em Bad} can be empty as approximation may add transitions that correspond to multiple steps. An important detail is to record the reason for divergence. If the set of {\em Bad} states is empty, divergence is due to approximation (line 10), $APPR$\footnote{Note that an over-approximation is computed for the negation of the formula, which gets propagated to each temporal operator.}. Otherwise, it is due to abstraction, in which case the {\em Deadend} and the {\em Bad} states are recorded as the conflicting states (line 13) to be passed to Craig Interpolation procedure as shown in Figure \ref{alg:cegaar}.

\begin{figure}[th!]
\begin{footnotesize}
\begin{algorithmic}[1]
\State {\bf CheckValidityAndDivergence}($s$: $S$, $s_{prev}$: $S$,$s^{\sharp}$: $S^{\sharp}$, $s^{\sharp}_{prev}$: $S^{\sharp}$, $f: ECTL$, $iter: ECTL \to \text{list of }\mathcal{P}(S^{\sharp})$, $witness: \text{list of }\mathcal{P}(S^{\sharp})$)
\State {\em Output:} $divType : \{ABST, APPR,NODIV,INVALID\}$, $divDepth: int$, $Conf: S \times S$   
\State $iter_f \gets iter(f)$
  \State $s^{\sharp}_{w} \gets s^{\sharp} \wedge iter_f.last()$
    \If{$s^{\sharp} = false$}
      {\bf return} $(INVALID,\_,\_)$
  \EndIf
   \State $witness.insert(model(s^{\sharp}_{w}))$
 \State $(dF,dT,Conf) \gets \text{\bf Divergence}(s_{X}, s, s^{\sharp}_{X}, s^{\sharp})$
     \If{$dF = true$}
        \State {\bf return}  $(dT,witness.size, Conf)$
     \Else 
         \State {\bf return} $(NODIV,\_,\_)$
    \EndIf     
\end{algorithmic}
\end{footnotesize}
\caption{Checks validity of the witness and divergence.}
\label{fig:checkvaldiv}
\end{figure}
   
\paragraph{Checking for Validity and Divergence} Figure \ref{fig:checkvaldiv} presents the algorithm that checks for validity of a witness for the given formula $f$ starting in the current abstract state by intersecting it with the solution set that corresponds to formula $f$ (line 4). If the abstract state $s^{\sharp}$ does not satisfy the formula $f$ then it returns $INVALID$ (line 5).  Otherwise, a model of the abstract state is inserted into the witness path (line 7).  It checks for divergence in the current states (line 8) and returns if one is found (line 10).  If not, it reports absence of divergence (line 12).

\begin{figure}[th!]
\begin{footnotesize}
\begin{algorithmic}[1]
\State {\bf GenAbsWit\_EX}($T=(S,I,R)$, $T^{\sharp}=(S^{\sharp}, I^{\sharp}, R^{\sharp})$, $s$: $S$, $s_{prev}$: $S$,$s^{\sharp}$: $S^{\sharp}$, $s^{\sharp}_{prev}$: $S^{\sharp}$, $f: ECTL$, $iter: ECTL \to \text{list of }\mathcal{P}(S^{\sharp})$, $witness: \text{list of }\mathcal{P}(S^{\sharp})$)
\State {\em Output:} $divType : \{ABST, APPR, NODIV,INVALID\}$, $divDepth: int$, $Conf: S \times S$   
  \State $(dT,dD,Conf) \gets \text{\bf CheckValidityAndDivergence}(s,s_{prev},s^{\sharp},s^{\sharp}_{prev},f,iter,witness)$  
  \If{$dT=INVALID$} \ {\bf return} $(dT,dD,Conf)$
  \ElsIf{$dT=NODIV$}
     \State $s^{\sharp}_X \gets post[R^{\sharp}](s^{\sharp})$
     \State $s_X \gets post[R](s)$
        \State {\bf return} $\text{\bf GenAbsWitHelper}(T, T^{\sharp}, s_{X}, s, s^{\sharp}_{X}, s^{\sharp}, f_1, iter, witness)$ 
  \Else  \ {\bf return} $(dT,dD,Conf)$      
  \EndIf   
\end{algorithmic}
\end{footnotesize}
\caption{An algorithm for generating an abstract witness for $EX$ formula up to a divergence point or up to the formula depth.}
\label{alg:witnessEX}
\end{figure}  
  
 \paragraph{Divergence in $EX$} Figure \ref{alg:witnessEX} presents the algorithm for generating an abstract witness for $EX$ formula up to a divergence point or up to the formula depth if no divergence is encountered. It first runs algorithm  {\bf CheckValidityAndDivergence}. If the abstract state satisfies the formula and does not have any divergence up to the current state, it checks for divergence 
 for subformula $f_1$ (line 8).
  
\begin{figure}[th!]
\begin{footnotesize}
\begin{algorithmic}[1]
\State {\bf GenAbsWit\_EU}($T=(S,I,R)$, $T^{\sharp}=(S^{\sharp}, I^{\sharp}, R^{\sharp})$, $s$: $S$, $s_{prev}$: $S$,$s^{\sharp}$: $S^{\sharp}$, $s^{\sharp}_{prev}$: $S^{\sharp}$, $f: ECTL$, $iter: ECTL \to \text{list of }\mathcal{P}(S^{\sharp})$, $witness: \text{list of }\mathcal{P}(S^{\sharp})$)
\State {\em Output:} $divType : \{ABST, APPR, NODIV,INVALID\}$, $divDepth: int$, $Conf: S \times S$   
\State $iter_f \gets iter(f)$
\If{$\exists k. 1 \leq k \leq iter_f.size$ s.t. $s^{\sharp} \wedge iter_f[k]  \not = false$ AND ($k=1$ OR $s^{\sharp} \wedge iter_f[k-1] = false$)}
  \State $s^{\sharp} \gets s^{\sharp} \wedge iter_f[k]$
   \State $witness.insert(model(s^{\sharp}))$
   \For{$i: k-1 \text{ to } 0$}
          \State $(dF,dT,Conf) \gets \text{\bf Divergence}(s, s_{prev}, s^{\sharp}, s^{\sharp}_{prev})$ 
          \If{$dF = true$}
              \State {\bf return}  $(dT,witness.size, Conf)$
          \ElsIf{$s^{\sharp}  \rightarrow iter_f[1]$}    {\bf break}
         \ElsIf{$k>0$}          
             \State $(dT,dD,Conf) \gets \text{\bf GenAbsWitnessHelper}(T, T^{\sharp}, s, s_{prev}, s^{\sharp}, s^{\sharp}_{prev}, f_1, iter, witness)$ 
             \If{NOT($dT=NODIV$)} {\bf return} $(dT,dD,Conf)$
             \EndIf   
            \State $s^{\sharp}_{prev} \gets s^{\sharp}$
            \State $s^{\sharp} \gets post[R^{\sharp}](^{\sharp}) \wedge iter_f[i]$
            \State $witness.insert(model(s^{\sharp}))$
            \State $s_{prev} \gets s$
            \State $s \gets post[R](s)$
          \EndIf  
   \EndFor 
   \State {\bf return} $\text{\bf GenAbsWitnessHelper}(T, T^{\sharp}, s, s_{prev}, s^{\sharp}, s^{\sharp}_{prev}, f_2, iter, witness)$ 
\Else   \  {\bf return} \ $(INVALID,\_,\_)$
\EndIf   
\end{algorithmic}
\end{footnotesize}
\caption{An algorithm for generating an abstract witness for $EU$ formula up to a divergence point or up to the formula depth.}
\label{alg:witnesshelperEU}
\end{figure}

 \paragraph{Divergence in $EU$} Figure \ref{alg:witnesshelperEU} presents the algorithm for generating an abstract witness for  $EU$ formula up to a divergence point or up to the formula depth if no divergence is encountered. If the abstract state $s^{\sharp}$ does not satisfy the formula $f$ then it returns $INVALID$ (line 24). Otherwise, the length of the shortest path, $k-1$, that reaches $f_2$ (line 4) and abstract state that reaches $f_2$ in that number of steps is computed (line 5) and a model is inserted to the witness path (line 6). Then it enters a loop (line 7) that runs until it reaches the state that satisfy $f_2$ (line 11). Inside the loop,  it checks divergence on a witness path for subformula $f_1$ (line 13) and if a divergence is not encountered then  it computes the next abstract and concrete states using the respective post operators (lines 17 and 20) and the loop continues. If no divergence is detected until the loop exits, divergence is checked for subformula $f_2$.

\begin{figure}[th!]
\begin{footnotesize}
\begin{algorithmic}[1]
\State {\bf GenAbsWit\_EG}($T=(S,I,R)$, $T^{\sharp}=(S^{\sharp}, I^{\sharp}, R^{\sharp})$, $s$: $S$, $s_{prev}$: $S$,$s^{\sharp}$: $S^{\sharp}$, $s^{\sharp}_{prev}$: $S^{\sharp}$, $f: ECTL$, $iter: ECTL \to \text{list of }\mathcal{P}(S^{\sharp})$, $witness: \text{list of }\mathcal{P}(S^{\sharp})$)
\State {\em Output:} $divType : \{ABST, APPR, NODIV,INVALID\}$, $divDepth: int$, $Conf: S \times S$   
  \State $s^{\sharp} \gets s^{\sharp} \wedge iter_f.last()$
  \If{$s^{\sharp} = false$}
     {\bf return} $(INVALID,\_,\_)$
  \EndIf  
  \State $start \gets witness.size() + 1$
  \State $witness.insert(model(s^{\sharp}))$
  \State $visited \gets false$
  \State $i \gets 1$
  \State $continue \gets true$ 
  \While{$i \leq overApprBound$ AND $continue$}
          \State $(dF,dT,Conf) \gets \text{\bf Divergence}(s, s_{prev}, s^{\sharp}, s^{\sharp}_{prev})$ 
          \If{$dF = true$}
              \State {\bf return}  $(dT,witness.size, Conf)$
          \Else    
            \State $s^{\sharp}_{prev} \gets s^{\sharp}$
            \State $s^{\sharp} \gets post[R^{\sharp}](^{\sharp}) \wedge iter_f.last()$
            \State $witness.insert(model(s^{\sharp}))$
            \State $s_{prev} \gets s$
            \State $s \gets post[R](s)$
            \If{$s^{\sharp} \implies visited$}
               $continue \gets false$
           \EndIf
           \State $visited \gets visited \vee s^{\sharp}$    
          \EndIf                      
        \State $i \gets i+1$
  \EndWhile
  \If{$continue$} {\bf return} $(APPR,witness.size(),\_)$
  \Else 
  \For{$ start \leq i \leq witness.size()$}
    \State  $(divType, divDepth,Conf) \gets$ \State $\text{\bf GenAbsWitnessHelper}(T, T^{\sharp}, s, s_{prev}, s^{\sharp}, s^{\sharp}_{prev}, f_1, iter, witness)$ 
    \If{$divType \not = NODIV$}
       \State {\bf return} $(divType, divDepth,Conf)$
    \EndIf
  \EndFor
   \State {\bf return} $(divType, divDepth,Conf)$
 \EndIf   
\end{algorithmic}
\end{footnotesize}
\caption{An algorithm for generating an abstract witness for $EG$ formula up to a divergence point or up to the formula depth.}
\label{alg:witnesshelperEG}
\end{figure}  
  
 \paragraph{Divergence in $EG$} Figure \ref{alg:witnesshelperEG} presents the algorithm for generating an abstract witness for  $EG$ formula up to a divergence point or up to the formula depth if no divergence is encountered. If the abstract state $s^{\sharp}$ does not satisfy the formula $f$ then it returns $INVALID$ (line 4).  Otherwise, a model of the abstract state is inserted into the witness path (line 6). Then it enters a loop until it either reaches a divergence (line 11), a cycle (line 20), or handles all the steps in the iterations by computing successors using the respective post operators (lines 16 and 19). If it does not find a cycle then it reports imprecision due to over approximation (line 26). Otherwise, it checks for divergence on the  witness path for subformula $f_1$ and returns the result if there is a divergence or invalidity (lines 32-34).  If no problem is encountered, it returns it after going through every state for the witness path for $EG$ (line 36).
 
\begin{figure}[th!]
\begin{footnotesize}
\begin{algorithmic}[1]
\State {\bf GenAbsWit\_AND}($T=(S,I,R)$, $T^{\sharp}=(S^{\sharp}, I^{\sharp}, R^{\sharp})$, $s$: $S$, $s_{prev}$: $S$,$s^{\sharp}$: $S^{\sharp}$, $s^{\sharp}_{prev}$: $S^{\sharp}$, $f: ECTL$, $iter: ECTL \to \text{list of }\mathcal{P}(S^{\sharp})$, $witness: \text{list of }\mathcal{P}(S^{\sharp})$)
\State {\em Output:} $divType : \{ABST, APPR, NODIV,INVALID\}$, $divDepth: int$, $Conf: S \times S$     
  \State $(dT,dD,Conf) \gets \text{\bf CheckValidityAndDivergence}(s,s_{prev},s^{\sharp},s^{\sharp}_{prev},f,iter,witness)$  
  \If{$dT=INVALID$} \ {\bf return} $(dT,dD,Conf)$
  \ElsIf{$dT=NODIV$}
         \State $(dT, dD, Conf) \gets \text{\bf GenAbsWitnessHelper}(T, T^{\sharp}, s, s_{prev}, s^{\sharp}, s^{\sharp}_{prev}, f_1, iter, witness)$  
         \If{$dT = NODIV$}
             \State $(dT, dD, Conf) \gets \text{\bf GenAbsWitnessHelper}(T, T^{\sharp}, s, s_{prev}, s^{\sharp}, s^{\sharp}_{prev}, f_2, iter, witness)$       
          \EndIf
          \State {\bf return} $(dT, dD, Conf)$
   \EndIf
\end{algorithmic}
\end{footnotesize}
\caption{An algorithm for generating an abstract witness for $AND$ formula up to a divergence point or up to the formula depth.}
\label{alg:witnesshelperAND}
\end{figure}

 \paragraph{Divergence in $AND$} Figure \ref{alg:witnesshelperAND} presents the algorithm for generating an abstract witness for  $AND$ formula up to a divergence point or up to the formula depth if no divergence is encountered. It first runs algorithm  {\bf CheckValidityAndDivergence}. If the abstract state satisfies the formula and does not have any divergence,
 it first checks for divergence in subformula $f_1$ (line 6). If it finds one, returns the details. Otherwise, checks for divergence in subformula $f_2$ (line 8) and returns whatever is found.

 \begin{figure}[th!]
\begin{footnotesize}
\begin{algorithmic}[1]
\State {\bf GenAbsWit\_OR}($T=(S,I,R)$, $T^{\sharp}=(S^{\sharp}, I^{\sharp}, R^{\sharp})$, $s$: $S$, $s_{prev}$: $S$,$s^{\sharp}$: $S^{\sharp}$, $s^{\sharp}_{prev}$: $S^{\sharp}$, $f: ECTL$, $iter: ECTL \to \text{list of }\mathcal{P}(S^{\sharp})$, $witness: \text{list of }\mathcal{P}(S^{\sharp})$)
\State {\em Output:} $divType : \{ABST, APPR, NODIV,INVALID\}$, $divDepth: int$, $Conf: S \times S$   
 \State $(dT,dD,Conf) \gets \text{\bf CheckValidityAndDivergence}(s,s_{prev},s^{\sharp},s^{\sharp}_{prev},f,iter,witness)$  
  \If{$dT=INVALID$} \ {\bf return} $(dT,dD,Conf)$
  \ElsIf{$dT=NODIV$}
         \State $(dT_1, dD_1, Conf_1) \gets \text{\bf GenAbsWitnessHelper}(T, T^{\sharp}, s, s_{prev}, s^{\sharp}, s^{\sharp}_{prev}, f_1, iter, witness)$  
         \If{$dT_1=INVALID$}
             \State {\bf return} $ \text{\bf GenAbsWitnessHelper}(T, T^{\sharp}, s, s_{prev}, s^{\sharp}, s^{\sharp}_{prev}, f_2, iter, witness)$       
         \Else     
              \State {\bf return} $(dT_1, dD_1, Conf_1)$             
          \EndIf
   \EndIf
\end{algorithmic}
\end{footnotesize}
\caption{An algorithm for generating an abstract witness for $OR$ formula up to a divergence point or up to the formula depth.}
\label{alg:witnesshelperOR}
\end{figure}

 \paragraph{Divergence in $OR$} Figure \ref{alg:witnesshelperOR} presents the algorithm for generating an abstract witness for  $OR$ formula up to a divergence point or up to the formula depth if no divergence is encountered. It first runs algorithm  {\bf CheckValidityAndDivergence}. If the abstract state satisfies the formula and does not have any divergence,
 it first checks for divergence in subformula $f_1$ (line 6). Unlike in the case of an $AND$ operator, there is a possibility of receiving an $INVALID$ result as the formula may not be satisfied in both subformulas. So if if it is not satisfied, it checks for a valid witness path or a divergence for subformula $f_2$ (line 8). 
  
\ignoreme{
\begin{figure}[th!]
\begin{footnotesize}
\begin{algorithmic}[1]
\State {\bf GenAbsWit\_ATOMIC}($T=(S,I,R)$, $T^{\sharp}=(S^{\sharp}, I^{\sharp}, R^{\sharp})$, $s$: $S$, $s_{prev}$: $S$,$s^{\sharp}$: $S^{\sharp}$, $s^{\sharp}_{prev}$: $S^{\sharp}$, $f: ECTL$, $iter: ECTL \to \text{list of }\mathcal{P}(S^{\sharp})$, $witness: \text{list of }\mathcal{P}(S^{\sharp})$)
\State {\em Output:} $divType : \{ABST, OVER\_APPR, UNDER\_APPR,NODIV,INVALID\}$, $divDepth: int$, $Conf: S \times S$   
  \State {\bf return} \  {\bf CheckValidityAndDivergence}$(s,s_{prev},s^{\sharp},s^{\sharp}_{prev},f,iter,witness)$  
 \end{algorithmic}
\end{footnotesize}
\caption{An algorithm for generating an abstract witness for $ATOMIC$ formula up to a divergence point or up to the formula depth.}
\label{alg:witnesshelperATOMIC}
\end{figure}

 \paragraph{Divergence in $ATOMIC$} Figure \ref{alg:witnesshelperATOMIC} checks the validity of the witness and if so, checks for divergence and returns the result.  
 }

\section{Experiments}
\label{sec:exp}

\begin{table*}[th!]
\centering
\begin{footnotesize}
\begin{tabular}{|r|r|r|r|r|r|}  \hline
{\bf Problem} & {\bf $|V|$} & \multicolumn{4}{c|}{\bf Trans. Rel.} \\ \cline{3-6}
{\bf Instance} &  &  {\bf BDD} & {\bf Poly} & {\bf (G)EQ} & {\bf \#Dis} \\ \hline
ticket2 & 4I, 4B & 65 & 3 & 4 & 3  \\ \hline
ticket4 & 6I, 8B &  195 & 5  & 16  & 5  \\ \hline
bakery2  & 2I, 4B & 54 & 1 & 0 & 1 \\ \hline
bakery4  & 4I, 8B & 260 & 17 & 54 & 9 \\ \hline
airportSM-tic2 & 9I, 8B&527 & 10 & 75 & 10 \\ \hline
airportSM-tic4 & 11I, 16B &  1236 & 9 & 57 & 9 \\ \hline
airportSM-bk2 & 7I , 8B &  542 & 13 & 73 & 11 \\ \hline
airportSM-bk4 & 9I, 16B&  1298 & 23 & 163 & 15 \\ \hline
charDriver2 & 4I, 11B&  1097 & 13 & 32 & 13 \\ \hline
charDriver4 & 4I, 21B & 1476 & 13 & 32 & 13 \\ \hline
\end{tabular}
\end{footnotesize}
\caption{Sizes of the problem instances. $|V|$, {\bf BDD}, {\bf Poly}, {\bf (G)EQ}, and {\bf \#Dis} represent number and types of state variables ($I$ for integer and $B$ for boolean), the size of the BDD, number of polyhedra, total number of equality and inequality constraints, and the number of disjuncts in the respective constraint, respectively.}
\label{table:size}
\end{table*}

We have applied the CEGAAR technique to various problems. The experiments have been executed on a 64-bit Intel Xeon(R) CPU with 8 GB RAM running Ubuntu 14.04 LTS. 
We have used two mutual exclusion protocols as the small benchmarks, the ticket algorithm \cite{And91} and Lamport's Bakery algorithm, and three larger benchmarks, which consisted two versions of the Airport Ground Network Traffic Control (AGNTC) model and a character special device driver model. AGNTC is a resource sharing model for multiple processes, where the resources are taxiways and runways of an airport ground network and the processes are the arriving and departing airplanes. We changed the AGNTC model given in \cite{YB09} to obtain two variants  by 1) using one of the two mutual exclusion algorithms, the ticket algorithm \cite{And91} ({\tt airportSM-tic}) and Lamport's Bakery algorithm ({\tt airportSM-bk}), for synchronization on one of the taxiways, 2) making parked arriving airplanes fly and come back to faithfully include the mutual exclusion model, i.e., processes go back to $think$ state after they are done with the critical section in order to attempt to enter the critical section again, and 3) removing the departing airplanes. 
The character-special device driver is a pedagogical artifact from a graduate level course. It models two modes, where one of the modes allows an arbitrary number of processes to perform file operations concurrently whereas the other mode allows only one 
process at a time. The synchronization is performed using semaphores that are modeled with integer variables. The {\tt ioctl} function is used to change from one mode to another when there are no other processes working in the current mode. The total number of processes in each mode is kept track of to transition from one mode to another in a safe way. The update of the process counters are also achieved using semaphores modeled with integer variables. We denote safety properties with suffix $-S$ and liveness properties with suffix $-L$. 

Table \ref{table:size} shows sizes of the problem instances in terms of integer $I$ and boolean $B$ variables and sizes of the state space, the initial state, and the transition relation, which were demonstrated in terms of he Binary Decision Diagram (BDD) size for the boolean domain, number of polyhedra and number of integer constraints for the integer domain, and number of composite\footnote{A composite formula consists of conjunction of a boolean and integer formula, where the former is represented with a BDD and the latter is represented using polyhedra.} disjuncts. 
ALV applies a simplification heuristic \cite{YB02c} to reduce size of the constraints and the data in Table  \ref{table:size} represents the values after simplification.

\begin{table}[th!]
\centering
\begin{footnotesize}
\begin{tabular}{|r|r|r|r||r|r|r|r|r|} \hline
{\bf Problem}  & {\bf FL} & \multicolumn{2}{c||}{\bf POLY} & \multicolumn{5}{c|}{\bf CEGAAR}  \\ \cline{2-9}
{\bf  Instance}& & {\bf Mem.} & {\bf VT} & {\bf Mem.} & {\bf VT} & {\bf RT} & {\bf \#R} & {\bf P?}   \\ \hline
ticket2-S & $A$ & 1.70 & 0.06 & \multicolumn{5}{c|}{$TO$}  \\ \hline
ticket2-S & $A$, $F$ & 1.30 & 0.01 & \multicolumn{5}{c|}{$TO$} \\ \hline
ticket2-L & $A$, $F$ & 1.42 & 0.02 & & & & & \\ \hline
bakery2-S & $A$ & 1.43 & 0.01 & 12.59 & 0.04 & 0.20 & 4 ABS & No  \\ \hline
bakery2-S& $A$, $F$ & 1.50 & 0.03 & 12.19 & 0.05 & 0.16 & 4 ABS & No \\ \hline
bakery4-S & $A$ & 25.69 & 19.39 &\multicolumn{5}{c|}{$TO$} \\ \hline
bakery4-S & $A$, $F$ &  \multicolumn{2}{c||}{$(\epsilon)$} & \multicolumn{5}{c|}{$TO$}  \\ \hline
bakery2-L & $A$, $F$ & 1.76 & 0.06 & 4.10 & 0.03 & 0.01 & 1 ABS & No  \\ \hline
bakery4-L & $A$, $F$ &   \multicolumn{2}{c||}{$(\epsilon)$} &  \multicolumn{5}{c|}{$OOM$}  \\ \hline
airportSM-bk2-S & $A$ & 6.47 & 1.45 & 110.40  & 9.69 & 2.59 & 7 ABS & Yes  \\ \hline
airportSM-bk2-S & $A$, $F$ & 6.07 & 0.57 & 83.18 & 30.90 & 0.35 & 1 ABS & Yes  \\ 
& & & & & & & 1 APP & \\
\hline
airportSM-bk4-S & $A$ &  \multicolumn{2}{c||}{$TO$}  & \multicolumn{5}{c|}{$TO$}  \\ \hline
airportSM-bk4-S & $A$, $F$ & 60.13 & 162.70 & \multicolumn{5}{c|}{$TO$} \\ \hline
airportSM-bk2-L & $A$, $F$ & 10.23 & 1.61 & 596.86 & 13.26 & 4.78 & 3 ABS & Yes  \\ \hline
airportSM-bk4-L & $A$, $F$ &   \multicolumn{2}{c||}{$TO$}  & \multicolumn{5}{c|}{$TO$}  \\ \hline
airportSM-tic2-S & $A$  & \multicolumn{2}{c||}{$UV$}  & 55.74 & 4.09 & 2.70 & 7 ABS & Yes  \\ \hline
airportSM-tic2-S & $A$,$F$ & 5.4 & 0.58 & 154.59 & 42.19  & 2.75 & 7 ABS & Yes \\ \hline
airportSM-tic4-S & $A$ &  \multicolumn{2}{c||}{$TO$}  &\multicolumn{5}{c|}{$TO$} \\ \hline
airportSM-tic4-S & $A$, $F$ & 36.20 & 98.76 & \multicolumn{5}{c|}{$TO$} \\ \hline
airportSM-tic2-L & $A$, $F$ & 9.79 & 1.18 & 45.32 & 15.71 & 0.01 & 1 APP & Yes \\ \hline
airportSM-tic4-L & $A$, $F$ &  \multicolumn{2}{c||}{$TO$} & \multicolumn{5}{c|}{$TO$}   \\ \hline
chardrv2-S & $A$ &  \multicolumn{2}{c||}{$TO$}   &  \multicolumn{5}{c||}{$TO$}  \\ \hline
chardrv2-S & $A$, $F$ & 1.49 & 14.40  &  \multicolumn{5}{c|}{$TO$}  \\ \hline
chardrv4-S & $A$ & \multicolumn{2}{c||}{$TO$}  & \multicolumn{5}{c|}{$TO$}  \\ \hline
chardrv4-S & $A$,$F$ &  \multicolumn{2}{c||}{$TO$}  & \multicolumn{5}{c|}{$TO$}  \\ \hline
\end{tabular}
\end{footnotesize}
\caption{Comparison of pure fixpoint approximation approach with CEGAAR for Partial Predicate Abstraction. $\epsilon$ means internal error in Omega Library. $TO$ means timeout (40 mins or more). $OOM$ means out of memory. $UV$ means unable to verify.}
\label{table:polyvscegaar}
\end{table}

\begin{table}[th!]
\centering
\begin{footnotesize}
\begin{tabular}{|r|r|r|r||r|r|} \hline
{\bf Problem}  & {\bf FL} & \multicolumn{2}{c||}{\bf POLY} & \multicolumn{2}{c|}{\bf PARTIAL PRED. ABS.} \\ \cline{3-6}
{\bf  Instance}& & {\bf Mem.} & {\bf VT} & {\bf Mem.} & {\bf VT}  \\ \hline
bakery2-S & $A$ & 1.43 & 0.01 & 5.52 & 0.06 \\ \hline
bakery2-L & $A$, $F$ & 1.76 & 0.06 & 3.02 & 0.03 \\ \hline
airportSM-bk2-S & $A$ & 6.47 & 1.45 &  14.40 &  0.49  \\ \hline
airportSM-bk4-S & $A$ &  \multicolumn{2}{c||}{$TO$}  & 34.09  &  9.27 \\ \hline
airportSM-bk2-L & $A$, $F$ & 10.23 & 1.61 & 20.15  & 2.27 \\ \hline
airportSM-tic2-S & $A$,$F$ & 5.4 & 0.58 & 10.76 & 2.32 \\ \hline
airportSM-tic4-S & $A$,$F$ & 36.20 &199.49 & 34.77 & 3.02 \\ \hline
airportSM-tic2-L & $A$, $F$ & 9.79 & 1.18 & 20.66 & 2.77   \\ \hline
chardrv2-S & $A$, $F$ & 1.49 & 14.40  & 35.52 & 1.94 \\ \hline
chardrv4-S & $A$,$F$ &  \multicolumn{2}{c||}{$TO$}  & 83.21 & 20.23 \\ \hline
\end{tabular}
\end{footnotesize}
\caption{Comparison of pure fixpoint approximation approach with Partial Predicate Abstraction when the seed set of predicates does not require refinement. $TO$ means timeout (40 mins or more). }
\label{table:polyvsppa}
\end{table}

Table \ref{table:polyvscegaar} shows verification results for pure polyhedra based analysis, POLY,  (no predicate abstraction) and for partial predicate abstraction with CEGAAR,  for which we used seed predicate sets that did not provide a conclusive result without refinement. We report the memory usage in MB, verification time, {\bf VT}, and refinement time, ${\bf RT}$, in secs, number of refinement steps, {\bf \# R.} (ABS means abstraction refinement and APP means approximation refinement), whether final refinement abstracted all integer variables ({\bf P?} being false). {\bf FL} denotes the flags enabled for ALV: $A$ denotes approximate fixpoint computation and $F$ denotes approximate reachable states computation. When $F$ is enabled an over-approximation of the reachable states is used to constrain the fixpoint solutions. 
We have used a timeout, $TO$, of 40 mins. 
As the results suggest CEGAAR is effective for small sizes of the instances with an exception for {\tt chardrv} benchmark. Those instances of CEGAAR that finished without a timeout, it took up to  7 abstraction refinements. Two of the instances, {\tt airportSM-bk2-s} and {\tt airportSM-tic2-L}, featured approximation refinements, which made the approximate reachable state computation more precise for the abstract transition system. Although CEGAAR could not compete with POLY, this is expected as the verification stage is performed from scratch after every refinement. 
As the data suggest, for the problem instances for which CEGAAR was effective, most of the time has been spent in the verification stage for large problems, {\tt airportSM-bk2} and {\tt airportSM-tic2}, and in the refinement stage for the smaller problem, {\tt bakery2-S}. 

In \cite{Yav16}, we demonstrated that partial predicate abstraction provides tremendous improvements when the seed predicates provide the necessary precision, i.e., no refinement is needed. Table \ref{table:polyvsppa} shows performance of partial predicate abstraction without refinement for the benchmarks for which such seed predicates could be identified. We basically used the predicate sets found by the CEGAAR technique and added extra predicates that did not reduce precision. As the data suggests the improvement has been in terms of 
verification time and the amount of improvement varied across the benchmarks and was greater for larger instances of the benchmarks. 

\begin{table}
\begin{footnotesize}
\begin{tabular}{|l|l|l|l|} \hline
{\bf Problem }  & {\bf Property} & {\bf Seed Pred. Set} & {\bf Final Pred. Set} \\ 
 {\bf Instance} & & & \\ \hline
bakery2-S & $\text{AG(!(p1=cs and pc=cs))}$ & $\{a<b\}$ & $\{a<b,b\geq1, $ \\
 & & & $42+44a \geq 43b,\}$ \\
 & & &  $42a+43b \leq 42, $ \\
 & & & $42b \leq a\}$ \\ \hline 
 bakery2-L & $\text{AG(p1=try implies }$ & $\{a<b\}$ & $\{a<b, b\geq 1\}$ \\
   & $\text{         AF(p1=cs))}$& & \\ \hline  
airportSM-bk2-S & $AG(numC3 \leq 1)$ & $\{numC3 \leq 1\}$ &$\{numC3 \leq 1, $ \\
& & & $numC3=0\}$ \\ \hline
 airportSM-bk2-L & $\text{AG(apc1=touchDownTry}$ & $\{numRW16R = 0,$ & $\{numRW16R = 0,$\\
 & $\text{     implies }$& $numB2A = 0\}$ &  $numB2A = 0,$\\
 & $\text{AF(apc1=touchDownCs))}$& & $numB2A = 1\}$\\ \hline
airportSM-tic2-S & $AG(numC3 \leq 1)$ & $\{numC3 \leq 1\}$ &$\{numC3 \leq 1, $ \\ 
& & &  $numC3=0\}$\\ \hline
 airportSM-tic2-L & $\text{AG(apc1=touchDownTry}$ & $\{numRW16R = 0,$ & $\{numRW16R = 0,$\\
 & $\text{     implies }$& $numB2A \geq 0\}$ &  $numB2A \geq 0\}$\\
 & $\text{AF(apc1=touchDownCs))}$& & \\ 
\hline 
\end{tabular}
\end{footnotesize}
\caption{The set of final predicates found by the CEGAAR technique for verifying an ACTL property for  given set of seed predicates.}
\label{table:polyvscegaar}
\end{table}

Table \ref{table:polyvscegaar} shows the verification instances for which the CEGAAR technique was successful in finding a set of predicates in the given time bound. The final predicate sets found did not differ whether we used the $A$ option or both $A$ and $F$ option. However, as it can be seen from  Table \ref{table:polyvscegaar}, these two settings had impact on the verification time, triggering an approximation refinement, and achieving a conclusive result.

\ignoreme{
\begin{table}
\begin{tabular}{|l|l|}
{\bf Problem Instance}  & {\bf Notes}  \\
barber &  \\
ticket2 &  \\
bakery2Safety & property safety, seed: $a < b$, infers $a=0, b=0$ , BEAUTY mode also verifies albeit finding different predicates\\
bakery2Liveness & property liveness, seed: $a < b$, infers $b=0$, $b=1$ , BEAUTY mode also verifies albeit finding different predicate $b>0$\\
airport2 &  \\
airport-bk2 &  property $numC3 \leq 1$, seed $numC3 \leq 1$, infers $numC3 = 0$, BEAUTY mode also verifies by finding the same predicate \\
airport-bk2Liveness &  property liveness, those seeds that are imprecise cause divergence due to approximation, e.g., a=b AND numRW16R > 1\\
airport-tic2 &  property safety $numC3 \leq 1$, seed $numC3 \leq 1$, infers $numC3 = 0$, BEAUTY mode also verifies by finding the same predicate \\
airport-tic2Liveness &  property liveness, those seeds that are imprecise cause divergence due to approximation, e.g., num16RWR=0 AND s=0 \\
chardrv - p1& p1/pred0.txt verifies \\
chardrv - p2 & 
\end{tabular}
\end{table}

\begin{table}
\begin{tabular}{|l|l|l|l|}
{\bf Problem Instance}  & {\bf Seed Predicate Set} &  \multicolumn{2}{c|}{\bf CEGAR-PA}  \\
chosen one & & & \\
\end{tabular}
\end{table}
}
\section{Related Work}
\label{sec:relwork}

\paragraph{Partial Predicate Abstraction} In \cite{GC10} a new abstract domain that combines predicate abstraction with numerical abstraction is presented. The idea is to improve precision of the analysis when the predicates involve numeric variables that are represented by an abstract numeric domain, e.g., a predicate on an array cell, where the domain of the index variables are represented using polyhedral domain. In our approach the predicates and the numeric variables do not have any interference. Although \cite{GC10} has evaluated the combined approach in the context of software model checking, our evaluation in the context of CTL model checking shows similar improvement in performance over complete numeric representation.

In \cite{JM06} Jhala et al. point out the inadequacy of generating predicates for integer domain based on weakest preconditions over counter-examples. They propose a complete technique for finding effective predicates when the system satisfies the property and involves a bounded number of iterations. 
The technique limits the range of constants to be considered at each refinement stage and avoids generation of diverging predicates in the interpolation stage by discovering new constraints that relate program variables. Unlike the examples considered in \cite{JM06}, the presented mutual exclusion algorithm does not have a bound and, therefore, it is not obvious whether that technique would be successful on ticket-like models. 

 Transition predicate abstraction \cite{PR07} is a technique that overcomes the inherent imprecision of state-based  predicate abstraction with respect to proving liveness properties. Although we also consider primed versions of variables in the predicate as part of the transition, our approach cannot handle predicates that directly relate primed and unprimed variables. In \cite{PR07}  such predicates can be handled as it uses the abstract transitions to label nodes of the abstract program. However, our approach is able to handle liveness properties that relate abstracted and concrete variables.

\paragraph{CEGAAR} There has been various approaches to improving performance of CEGAR in the context of CTL/LTL model checking by   
a combination of machine learning and linear integer programming to find a compact refinement predicate \cite{CGK02}, 
monitoring SAT checking phase to identify variables relevant to the refinement \cite{CCK02}
eliminating redundant predicates \cite{CGT03}, 
inferring compact refinement predicates using the conflict graphs for a SAT query \cite{CTV03}, and 
extracting clauses used in proof of unsatisfiability (proof of absence of a counter-example up to a depth) \cite{MA03}. 
Our approach finds the minimum set of predicates that would make the analysis precise using a BFS search strategy. Additionally, as a novel approach, it supports abstraction refinement with approximation refinement. 

\section{Conclusion}
\label{sec:conc}

\ignoreme{
State-explosion is an inherent problem in model checking. The abstract interpretation framework provides a systematic way of combining abstractions and approximations to push the limits of automated verification, specifically model checking. In this paper we have combined two  techniques, predicate abstraction and approximate fixpoint computations, that have been effective in the verification of infinite-state systems. This hybrid approach proves to be advantageous when the fixpoint computations do not converge for a given correctness property. In such cases letting part of the state space that causes non-convergence to be handled by fixpoint approximations and the rest of the state space abstracted using predicate abstraction can improve the verification time and in some cases the memory requirements. 
}

We have presented a counter-example guided abstraction and approximation refinement (CEGAAR) for partial predicate abstraction. 
As a hybrid approach, partial predicate abstraction  combines predicate abstraction with fixpoint approximations so that when approximate fixpoint computation is more effective than predicate abstraction in terms of providing the necessary precision, the state explosion can be dealt with the help of predicate abstraction. To deal with verification configurations that are not precise enough for getting a conclusive enough, CEGAAR identifies the source of imprecision and applies the relevant refinement technique.  
We have implemented the proposed approach 
 in the context of Action Language Verifier, an infinite-state symbolic model checker that performs approximate CTL model checking. 
 Experimental results show that approximation refinement and abstraction refinement can work harmoniously to achieve a conclusive result. 
 Although partial predicate abstraction provides significant improvements for both safety and liveness verification,  CEGAAR does not scale to large problems. 
 For future work, we would like to incorporate predicate selection  heuristics to scale CEGAAR to larger problems.

\clearpage 



\bibliographystyle{elsarticle-num} 
\bibliography{jlamp17}





\end{document}